\documentstyle[sprocl,epsf]{article}

\def\Journal#1#2#3#4{{#1} {\bf #2}, #3 (#4)}

\def\SCN{\em Science}

\def\PRL{\em Phys. Rev. Lett.}

\def\PRB{{\em Phys. Rev.} B}

\def\be{\begin{equation}}
\def\ee{\end{equation}}
\def\bea{\begin{eqnarray}}
\def\eea{\end{eqnarray}}
\begin{document}

\title{HIGH TEMPERATURE RELAXATIONAL DYNAMICS\\
 IN \\
 LOW-DIMENSIONAL
QUANTUM FIELD THEORIES}

\author{Subir Sachdev}

\address{Department of Physics, Yale University\\
P.O. Box 208120,
New Haven, CT 06520-8120, USA\\
Email: subir.sachdev@yale.edu}


\maketitle\abstracts{
This paper presents a unified perspective on the results of two
recent works (C.~Buragohain and S.~Sachdev cond-mat/9811083
and S.~Sachdev cond-mat/9810399) along with additional background.
We describe the low frequency, non-zero temperature, order parameter
relaxational dynamics of a number of systems in the vicinity of
a quantum critical point. The dynamical correlations are
properties of the high temperature limit of
renormalizable quantum field theories in spatial dimensions
$d=1,2$.
We study, as a function of $d$ and the number of order parameter
components, $n$, the
crossover from the finite
frequency, ``amplitude fluctuation'', gapped quasiparticle mode in the quantum
paramagnet (or Mott insulator),
to the zero frequency ``phase'' ($n \geq 2$) or ``domain wall'' ($n=1$)
relaxation mode
near the ordered phase. Implications for dynamic measurements on
the high temperature superconductors and antiferromagnetic spin chains
are noted.
}

\begin{center}
\tt
Published in \\
{\it Highlights in Condensed Matter Physics}, \\
APCTP/ICTP Joint International Conference,\\
 Seoul, Korea, June 12-16, 1998,\\
edited by B.~K.~Chung and M.~Virasoro \\
World Scientific, Singapore (2000). \\
Report No. cond-mat/9811110.
\end{center}

\section{Introduction}
\label{intro}

Recent neutron scattering experiments by Aeppli {\em et
al.}\cite{aeppliscience} have studied the
two-dimensional incommensurate spin correlations in
the normal state of the high
temperature superconductor ${\rm La}_x {\rm Sr}_{1-x} {\rm Cu O}_4$
at $x\approx 0.15$ in some detail.
In particular, they have determined the scattering cross section
over a significant portion of the wavevector, frequency and
temperature space. One of their striking observations has been
that, while the dynamic structure factor of the electronic spins
has quite an intricate functional form over this three-dimensional
parameter space, the results become quite simple and explicable
when interpreted in terms of the non-zero temperature dynamical
properties of a system in the vicinity of a quantum critical
point \cite{SY,CSY}.
The data are consistent with such an interpretation over
a decade in temperature and in over two decades of the peak
scattering amplitude, and indicate that there is a nearby
quantum critical point with dynamic critical exponent $z \approx 1$.
The nature of the measured spin correlations indicates that this
quantum critical point is the position of a quantum phase
transition to an insulating, charge- and spin-ordered ground state.
Varying the doping concentration, $x$, alone is not expected to be
sufficient to access this ordered state, and
authors \cite{aeppliscience} asserted that a second tuning
parameter (`$y$') is necessary; {\em e.g.} it is known that
replacing some of the ${\rm Sr}$
by ${\rm Nd}$ allows one to move along the $y$ axis \cite{tran}.

In this paper, we will review recent theoretical work studying
non-zero temperature ($T$)
dynamical response functions closely related to those measured in
the neutron scattering experiment of Aeppli {\em et
al.} \cite{aeppliscience}
We will examine a simple class of models in spatial dimension $d=1,2$, which exhibit
quantum ordering transitions with $z=1$. All of the models we
shall study are closely related to the following quantum field
theory of a real, $n$-component, scalar field $\phi_{\alpha}$
($\alpha = 1 \ldots n$):
\begin{eqnarray}
&& {\cal Z}_Q = \int {\cal D} \phi_{\alpha} (x, \tau) \exp \left( - \int d^d x
\int_0^{1/T} d \tau \, {\cal L}_Q
\right) \nonumber \\
&& {\cal L}_Q =  \frac{1}{2} \left[
 \frac{1}{c^2} (\partial_{\tau} \phi_{\alpha})^2 + (\nabla_x \phi_{\alpha})^2 +
(r_c + r) \phi_{\alpha}^2 \right]
+ \frac{u}{4!} \left( \phi_{\alpha}^2 \right)^2 .
\label{calsq}
\end{eqnarray}
Here we are using units with $\hbar = k_B = 1$,
$x$ is the $d$-dimensional spatial co-ordinate, $\tau$ is
imaginary time, $c$ is a velocity, and $r_c$, $r$ and $u$
are coupling constants.
The co-efficient of the $\phi_{\alpha}^2$ term (the `mass' term)
has been written as
$r + r_c$ for convenience.
The quartic non-linearity $u$ makes ${\cal Z}_Q$ as interacting
quantum field theory, and it ultimately responsible for the
dynamic relaxation phenomena we shall describe.
For appropriate values of $d$ and $n$,
the model ${\cal Z}_Q$ exhibits a $T=0$ quantum transition between
an ordered phase with $\langle \phi_{\alpha} \rangle \neq 0$, and a quantum
paramagnet with complete ${\rm O}(n)$ symmetry; the value of $r_c$
is chosen so that this transition occurs at $r=0$.
In physical applications of ${\cal Z}_Q$, the case $n=1$
describes the Ising model in a transverse field, the case $n=2$
case describes a superfluid-insulator transition (with
$ \phi_1 + i\phi_2$, the complex superfluid order parameter),
and $n=3$ describes spin fluctuations in a collinear quantum
antiferromagnet (with $\phi_{\alpha}$ the antiferromagnetic order
parameter).

We will be interested in the real-time dynamic properties of ${\cal Z}_Q$
in the high $T$ limit of the continuum quantum field theory (in
some cases, this is also the `quantum critical' region \cite{CHN}).
More specifically, in the vicinity of a quantum critical point,
the theory ${\cal Z}_Q$ can be characterized by two distinct
energy scales. The first, is a low energy scale, characterizing
the deviation from the system from the quantum critical point at
$r=0$: this varies as $b |r|^{z \nu}$, where $\nu$ is the
correlation length exponent, and $b$ is a non-universal,
cutoff-dependent constant needed to make the expression have
physical units of energy; this low energy scale is the central
quantity characterizing the continuum quantum field theory. The
second, is a high energy scale, and is of order $c \Lambda$, where
$\Lambda$ is a high-momentum cutoff needed at intermediate stages
to define the continuum limit of ${\cal Z}_Q$; this energy scale
plays no role in the continuum quantum field theory. We will be
assuming here that the temperature is {\em in between} these
two energy scales {\em i.e.}
\begin{equation}
b |r|^{z \nu} \ll T \ll c \Lambda.
\label{k1}
\end{equation}

We will characterize the response of the system by the dynamic
susceptibility, $\chi (k, \omega_n)$, defined by
\begin{equation}
\chi( k ,  \omega_n ) \equiv \frac{1}{n} \int_0^{1/T} d \tau \int d^d x \,
\sum_{\alpha=1}^{n}
\langle \phi_{\alpha} (x, \tau) \phi_{\alpha} (0,0) \rangle e^{-i(kx - \omega_n
\tau)},
\label{defchi}
\end{equation}
where
$k$ is the wavevector, $\omega_n$ the imaginary
frequency; throughout we will use the symbol $\omega_n$ to refer to
imaginary frequencies, while the use of
$\omega$ will imply the expression has been analytically continued to real
frequencies. Further, the static susceptibility, $\chi (k)$, is
defined by
\begin{equation}
\chi (k) \equiv \chi(k, \omega_n = 0).
\label{defstat}
\end{equation}
We shall be especially interested here in a {\em relaxation
function} ${\cal R} (k, \omega)$, which we define as
\begin{equation}
{\cal R}(k, \omega) = \frac{1}{\chi(k)} \frac{2}{\omega} \mbox{Im}
\chi (k, \omega).
\label{defrelax}
\end{equation}
This is an even function of $\omega$ with the dimensions of time, and the Kramers-Kronig
relation implies that
\begin{equation}
\int_{-\infty}^{\infty} \frac{d \omega}{2 \pi} \, {\cal R} (k,
\omega) = 1.
\label{k2}
\end{equation}
After a Fourier transform to real time, ${\cal R}(k, t)$ (with
${\cal R}(k, t=0) = 1$) describes
the time-dependent relaxation of spin correlations at wavevector $k$
due to quantum and thermal fluctuations.

(In a regime where the predominant fluctuations have an energy much
smaller than $T$, the low frequency dynamics becomes effectively
classical, and then the fluctuation-dissipation theorem implies
that
\begin{equation}
{\cal R} (k, \omega) \approx \frac{S(k, \omega)}{S(k)},
\label{k3}
\end{equation}
where $S(k, \omega)$ is the dynamic structure factor, and $S(k)$
is the equal-time structure factor.
However, this relationship is not generally true for a quantum system, and for clarity,
we will always use (\ref{defrelax}) as our defining relation.)

The following sections will describe the behavior of
${\cal R} (k, \omega) $ in the high $T$ limit of the continuum
quantum theory ${\cal Z}_Q$ for various physically interesting cases
in low dimensions: we will consider the case $n=1$, $d=1$ in
Section~\ref{sec:ising}, the case $n=3$, $d=1$ in
Section~\ref{sec:1drot}, and all values of $n$ in $d=2$
in Section~\ref{sec:eps}.

\section{Ising chain in a transverse field}
\label{sec:ising}

The Ising chain in a transverse field has the Hamiltonian
\begin{equation}
H_I = -J \sum_{i} \left(g \hat{\sigma}^x_i + \hat{\sigma}^z_{i} \hat{\sigma}^z_{i+1}
\right),
\label{ising1}
\end{equation}
where $\hat{\sigma}^{x,z}_i$ are Pauli matrices on the sites, $i$,
of a $d=1$ chain with lattice spacing $a$.
This model has a quantum critical point at
$g=1$, whose vicinity is believed to be described by the $d=1$,
$n=1$ case of ${\cal Z}_Q$. Under this mapping, the field-theoretic
correlators of $\phi_{\alpha}$ under ${\cal Z}_Q$ are equivalent
to the long-distance, long-time correlators of the order parameter
$\hat{\sigma}^z$ under the lattice model $H_I$.

\begin{figure}[t]
\epsfxsize=4in
\centerline{\epsffile{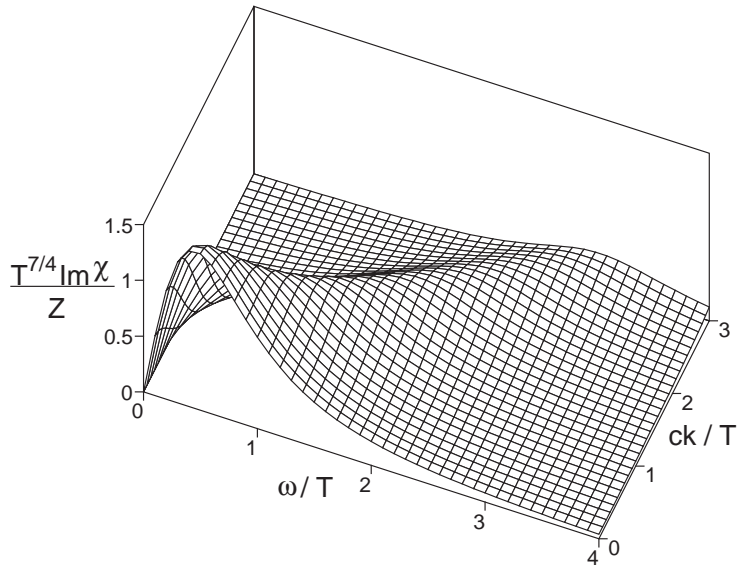}}
\caption{High temperature dynamic susceptibility, $ T^{7/4} \mbox{Im} \chi (k, \omega)/Z$ in
(\protect\ref{ising35}) of the quantum Ising chain ($d=1$, $n=1$)
as a function of $\omega/T$
and $ck/T$. }
\label{koreaf1}
\end{figure}

\begin{figure}[t]
\epsfxsize=4in
\centerline{\epsffile{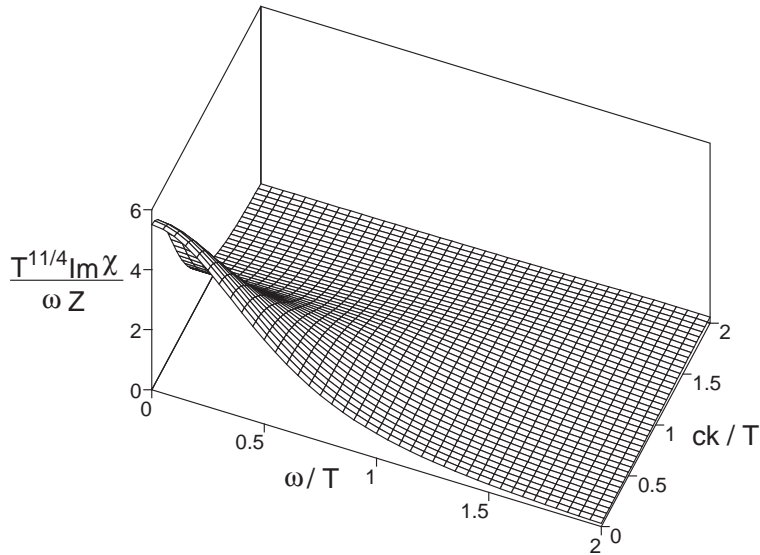}}
\caption{Plot of the spectral density of the quantum Ising chain
($d=1$, $n=1$)
$T^{11/4} \mbox{Im} \chi (k, \omega)/\omega Z$
as a function of $\omega/T$ and $c k /T$. Note that this is simply
the quantity in Fig~\protect\ref{koreaf1} divided by $\omega$.
The reactive peaks at $\omega \approx c k$ in Fig~\protect\ref{koreaf1}
are essentially invisible, and the plot is dominated by a large relaxational
peak at zero wavevector and frequency.}
\label{koreaf2}
\end{figure}

\begin{figure}[t]
\epsfxsize=4in
\centerline{\epsffile{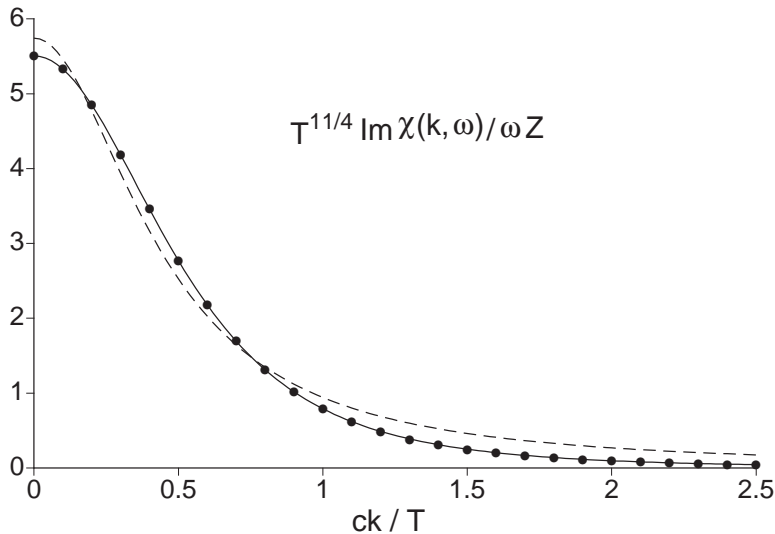}}
\caption{Comparison of the predictions of (\protect\ref{ising35})
(dots) and (\protect\ref{ising36z}) (solid line) for $\mbox{Im}\chi(k, \omega)/\omega$
of the quantum Ising chain ($d=1$, $n=1$)
at $\omega=0$ as a function of $c k/T$.
The best fit parameters in (\protect\ref{ising36y})
were used. The function (\protect\ref{ising36z}) yields the
{\em square} of a Lorentzian
as a function of $k$; a best fit by just a
Lorentzian
is also shown (dashed line), and is much poorer.}
\label{koreaf3}
\end{figure}

\begin{figure}
\epsfxsize=4in
\centerline{\epsffile{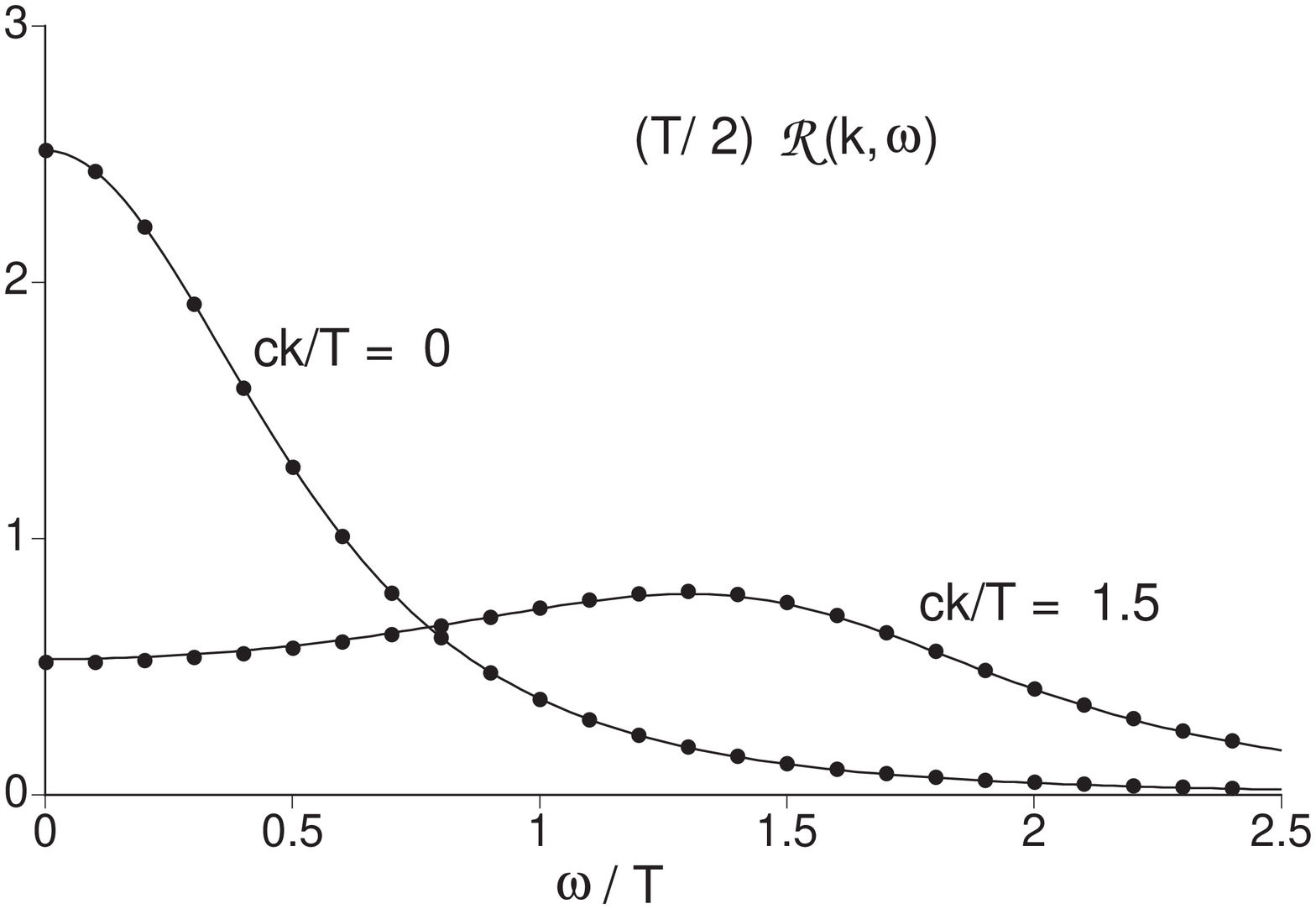}}
\caption{
Comparison of the predictions of (\protect\ref{ising35})
(dots) and (\protect\ref{ising36z}) (solid line) for
the relaxation function $(T/2)
{\cal R} (k, \omega)$ of the quantum Ising chain ($d=1$, $n=1$)
as a function of $\omega/T$
at $ck/T=0,1.5$.
}
\label{koreaf4}
\end{figure}

The continuum high $T$ dynamical response of (\ref{ising1}) can
be computed exactly by a special trick relying on the conformal
invariance of the critical theory at $g=1$ \cite{statphys}.
The result for $\chi (k, \omega)$ is
\begin{equation}
\chi (k , \omega ) = \frac{Z c}{T^{7/4}} \frac{G_I (0)}{4 \pi}
\frac{\Gamma(7/8)}{\Gamma(1/8)}
\frac{\displaystyle \Gamma \left( \frac{1}{16} - i \frac{\omega + c k}{4 \pi T} \right)
\Gamma \left( \frac{1}{16} - i \frac{\omega - c k}{4 \pi T} \right)}
{\displaystyle \Gamma \left( \frac{15}{16} - i \frac{\omega + c k}{4 \pi T} \right)
\Gamma \left( \frac{15}{16} - i \frac{\omega - c k}{4 \pi T} \right)}.
\label{ising35}
\end{equation}
Here $Z = J^{-1/4}$, $c=2 J a$, and $G_I (0) = 0.858715 \ldots$.
We show a plot of $\mbox{Im} \chi $ in Fig~\ref{koreaf1}.
For $\omega, ck \gg T$ there is a
well-defined `reactive' peak in $\mbox{Im} \chi $ at $\omega \approx c k$
reflecting the excitations of the $T=0$ quantum critical point,
which have an energy threshhold at $\omega = ck$.
However the low frequency dynamics is quite different, and
for $\omega, ck \ll T$ we cross-over to
the {\em quantum relaxational}
regime~\cite{CSY}.
This is made clear by an examination of $\mbox{Im} \chi (k, \omega)/\omega$
as a function of $\omega/T$ and $c k/T$, which is
shown in
Fig~\ref{koreaf2}.
Now the reactive peaks at $\omega \sim c k$ are just about
invisible, and the spectral density is dominated by a large
relaxational peak at zero frequency. We can understand the
structure of Fig~\ref{koreaf2} by expanding the inverse of
(\ref{ising35}) in powers of $k$ and $\omega$; this expansion has
the form
\begin{equation}
\chi(k, \omega) =
\frac{\chi (0)}{1 - i (\omega/\omega_1) + k^2 \widetilde{\xi}^2 -
(\omega/\omega_2)^2},
\label{ising36z}
\end{equation}
where $\chi(0)\sim T^{-7/4}$, and
$\omega_{1,2}$ and $\widetilde{\xi}$ are parameters
characterizing the expansion. For $k$ not too large,
the $\omega$ dependence in (\ref{ising36z}) is simply the
response of a strongly damped harmonic oscillator: this is the reason
we have identified the low frequency dynamics as ``relaxational''.
The function in (\ref{ising36z}) provides an excellent description
of the spectral response in Fig~\ref{koreaf2}. We determined the
best fit values of the parameters $\omega_{1,2}$ and $\widetilde{\xi}$
by minimizing the mean square difference between the values of $\mbox{Im}
\chi (k, \omega)/\omega$ given by (\ref{ising36z})
and (\ref{ising35}) over the range $0 < \omega < 2 T$ and $0 < c k < 2 T$
and obtained
\begin{eqnarray}
\omega_1 &=& 0.396~T\nonumber \\
\omega_2 &=& 0.795~ T\nonumber \\
\widetilde{\xi} &=& 1.280~c/T.
\label{ising36y}
\end{eqnarray}
The quality of the fit is shown in Figs~\ref{koreaf3} and~\ref{koreaf4}.
In Fig~\ref{koreaf3}
we compare the predictions of (\ref{ising35}) and (\ref{ising36z}) for
$\mbox{Im} \chi(k, \omega)/\omega$ at $\omega=0$ as a function of $ck/T$.
The form (\ref{ising36z}) predicts a Lorentzian-squared response
function and this is seen to provide a better fit than a
Lorentzian---a similar Lorentzian-squared response was used in
analyzing the data in Ref.~1.
In Fig~\ref{koreaf4} we plot the predictions of (\ref{ising35})
and (\ref{ising36z}) for ${\cal R} (k, \omega)$
at $ck/T=0,1.5$ as a function of $\omega/T$.
For $k=0$ ($\omega=0$) there is a large overdamped peak at $\omega=0$
($k=0$), but a weak reactive peak at $\omega \sim c k$ does make an
appearance at larger wavevectors or frequencies.

For an alternative, and more precise, characterization of the relaxational dynamics
we can introduce the relaxation rate $\Gamma_R$ defined by
\begin{equation}
\Gamma_R^{-1} \equiv \frac{{\cal R} (0,0)}{2};
\label{ising36}
\end{equation}
we have chosen this definition because
for the suggestive functional form (\ref{ising36z}), $\Gamma_R =
\omega_1$, the frequency characterizing the damping.
However, using (\ref{ising35}) we determine:
\begin{eqnarray}
\Gamma_R &=& \left( 2 \tan \frac{\pi}{16} \right) \frac{k_B T}{\hbar} \nonumber \\
&\approx& 0.397825 \frac{k_B T}{\hbar},
\label{ising37}
\end{eqnarray}
where we have inserted physical units to emphasize the
universality of the result.

An important property of the present high $T$
dynamical results is that
at the scale of the characteristic rate $\Gamma_R$,
the dynamics of the system involves intrinsic quantum effects
which cannot be neglected. Description by an effective classical
model would require that $\Gamma_R \ll k_B T / \hbar$,
which is thus not satisfied here.

To obtain an intuitive physical picture of the above
quantum relaxational dynamics, it is useful to consider
approaching the high $T$ limit by gradually rising the
temperature (while keeping all other couplings fixed)
from the low $T$ limit where $T \ll b |r|^{z \nu}$.

First, consider the low $T$ limit on the magnetically ordered
side. Here the excitations above the ground state are
`domain walls' which separate regions in which the Ising
order parameter has opposite signs. These domain walls can
move easily without significant change of energy, and their
low energy motion leads to a large relaxational peak in ${\cal R} (k, \omega)$
at $\omega=0$, $k=0$~\cite{apy}. At very low $T$, the domain walls
are very dilute, and their spacing is much larger than their
thermal de Broglie wavelengths---consequently their motion can be
described in a classical model. However, as $T$ is raised into the
high $T$ regime, their spacing becomes of order their de Broglie
wavelength, and the relaxation rate of their collisional dynamics
becomes of order $T$: this is leads to the relaxational peak in
Fig~\ref{koreaf4}.

Second, we can begin by considering the low $T$ limit on the
quantum paramagnetic side. Now the excitations are local
`flipped spins' which require a finite energy, $\Delta$, to create
them. So there is a sharp peak in ${\cal R} (k, \omega)$
at $\omega = \Delta$, which is broadened by collisions with the
dilute, classical gas of pre-existing quasiparticles.
In the language of the field $\phi_{\alpha}$, this finite
frequency peak arises from {\em amplitude fluctuations} in $\phi$
about a local minimum in its effective potential.
As $T$ is
raised, the quasiparticle gas becomes dense with the mean-particle spacing becoming
of order their de Broglie wavelength, the quasiparticle
line-width becomes of order $T$, and the peak in ${\cal R} (k, \omega) $ eventually
moves to $\omega = 0$.

\section{Gapped, Heisenberg antiferromagnetic chains}
\label{sec:1drot}

In this section we will consider ${\cal Z}_Q$ for the case $d=1$,
$n=3$, which describes the low energy fluctuations of certain
one-dimensional quantum antiferromagnets: chains of integer spin
$S$, or $p$-leg ladders with $p$ even.
The ground state of these systems is always a quantum
paramagnet with an energy gap, $\Delta$,
and there is no magnetically ordered state. So strictly speaking,
there is no quantum critical point, and it may appear that the general arguments of
Section~\ref{intro} do not apply. However, it is still possible to
define a universal, continuum high $T$ limit of the quantum
theory. We have to replace the condition (\ref{k1}) by
\begin{equation}
\Delta \ll T \ll c \Lambda \sim J,
\label{k4}
\end{equation}
where $J$ is a typical exchange constant of the antiferromagnet,
but, unlike (\ref{k1}), the energy gap, $\Delta$, is always non-zero.
So we have to pick a system with
$\Delta$ much smaller than any microscopic energy scale.
As $\Delta$ becomes exponentially small as $S$ (or $p$) is
increased, this is quite easily achieved. In a formal sense, we
can consider the following an analysis of the high $T$ dynamics
of the `quantum' critical point reached in the limit
$S \rightarrow \infty$, when $\Delta$ becomes vanishingly small.

The high $T$ limit introduced above has been analyzed in some
detail in two recent papers \cite{dslong,chiran}.
It was argued that there is an effective {\em classical non-linear
wave} problem that describes the long time relaxation dynamics.
The degrees of freedom of this classical model
are a 3-component unit length field ${\bf n} (x, t)$, ${\bf n}^2 = 1$,
which describes the local orientation of the antiferromagnetic
order parameter, and its canonically conjugate angular momentum,
${\bf L} (x, t)$, which measures the ferromagnetic component of the
local spin density. The equal time correlations of ${\bf L}$
and ${\bf n}$ are described by the following continuum
classical partition function
\begin{eqnarray}
&& {\cal Z}_{1C} =  \int {\cal D} {\bf n} (x) {\cal D} {\bf L} (x)
\delta( {\bf n}^2 - 1) \delta( {\bf L} \cdot {\bf n}) \exp\left(-\frac{{\cal H}_{1C}}{T}
\right) \nonumber \\
&& {\cal H}_{1C} = \frac{1}{2} \int d x \, \left[ T\xi
\left( \frac{d {\bf n} }{d x} \right)^2
+ \frac{1}{\chi_{u \perp}} {\bf L}^2\right].
\label{1drot44}
\end{eqnarray}
The parameters in this partition function are determined
universally in terms of $\Delta$, $T$, and $c$, and exact expressions
are known in the limit (\ref{k4}) \cite{dslong,chiran}. The correlation length, $\xi$,
is given by
\begin{equation}
\xi = \frac{c}{2 \pi T} \ln \left( \frac{32 \pi e^{-(1 + \gamma)}
T}{\Delta} \right),
\label{k5}
\end{equation}
where $\gamma$ is Euler's constant. The quantity $\chi_{u \perp}$
is the susceptibility to a uniform magnetic field (which couples
to the ferromagnetic moment) in a direction orthogonal to the
local antiferromagnetic order; it is related to the rotationally
averaged uniform susceptibility, $\chi_u$, by
\begin{equation}
\chi_{u} = \frac{2}{3} \chi_{u \perp},
\label{k6}
\end{equation}
and $\chi_u$ is given by
\begin{equation}
\chi_u = \frac{1}{3 \pi c} \ln \left( \frac{32 \pi e^{-(2 + \gamma)}
T}{\Delta} \right).
\label{k7}
\end{equation}

As we will see below, with these parameters in hand, the
characteristic
excitation of the classical model (\ref{1drot44}) has energy
$\varepsilon \sim T/\ln(T / \Delta)$. For $T \gg \Delta$, this is
parametrically smaller than $T$. So the occupation number of the
wave modes with energy $\varepsilon$ will be much larger than
unity, and the quantum Bose function will take the classical
equipartition value $T/\varepsilon$. This is the argument which
justifies use of a classical model in this high $T$ limit.

All equal time correlations of the model (\ref{1drot44}) can be
computed exactly. As this is a model to which
the classical fluctuation-dissipation theorem applies,
the equal time, two-point ${\bf n}$ correlator is directly related
to the static susceptibility; the underlying quantum fluctuations
however do induce an overall wavefunction renormalization
factor \cite{dslong,chiran}. The two point ${\bf n}$ correlator
decays exponentially on the scale $\xi$, and by its Fourier
transform to momentum space we obtain
\begin{equation}
T \chi (k) = {\cal A}
\left[ \ln \left(\frac{T}{\Delta} \right) \right]^2
\frac{2\xi/3}{(1 + k^2 \xi^2)}.
\label{1drotss2}
\end{equation}
Here ${\cal A}$ is a non-universal amplitude which determines the
scale of the field $\phi_{\alpha}$, and the multiplicative logarithmic factor
comes from the underlying quantum fluctuations; the remaining is
just the Fourier transform of $e^{-|x|/\xi}/3$, the $1/3$ coming
from the $1/n$ in (\ref{defchi}).

Let us now turn to the unequal time correlations. To obtain these,
we have to supplement ${\cal Z}_{1C}$ with equations of motion, which
have been argued \cite{dslong,chiran} to be the Hamilton-Jacobi
equations associated with the Poisson brackets of ${\bf n}$
and its canonically conjugate angular momentum ${\bf L}$:
\begin{eqnarray}
\left\{ L_{\alpha} (x) , L_{\beta} (x') \right\}_{PB} &=&
\epsilon_{\alpha \beta \gamma} L_{\gamma} (x) \delta(x-x') \nonumber \\
\left\{ L_{\alpha} (x) , n_{\beta} (x') \right\}_{PB} &=&
\epsilon_{\alpha \beta \gamma} n_{\gamma} (x) \delta(x-x') \nonumber \\
\left\{ n_{\alpha} (x) , n_{\beta} (x') \right\}_{PB} &=& 0.
\label{1drot45}
\end{eqnarray}
From this, and (\ref{1drot44}), we obtain directly the
equations of motion for the quasi-classical waves\index{quasi-classical!waves}
\begin{eqnarray}
\frac{\partial {\bf n}}{\partial t} &=& \{ {\bf n}, {\cal H}_{1C} \}_{PB}
\nonumber \\
&=& \frac{1}{\chi_{u \perp}} {\bf L} \times {\bf n}
\nonumber \\
\frac{\partial {\bf L}}{\partial t} &=& \{ {\bf L}, {\cal H}_{1C} \}_{PB}
\nonumber \\
&=& (T \xi) {\bf n} \times \frac{\partial^2 {\bf n}}{
\partial x^2}.
\label{1drot45a}
\end{eqnarray}
To compute the needed unequal time correlation functions,
pick a set of initial conditions for ${\bf n} (x)$, ${\bf L} (x)$ from the
ensemble (\ref{1drot44}). Evolve these deterministically in time using
the equations of motion (\ref{1drot45a}). The value of the correlator
is then the product of the appropriate time-dependent fields, averaged over
the set of all initial conditions.
We
also note here that simple analysis of the differential equations
(\ref{1drot45a}) shows that small disturbances about a nearly ordered ${\bf n}$
configuration travel with a characteristic velocity
$c(T)$ given by
\begin{equation}
c(T) = (T \xi (T) / \chi_{u \perp} (T) )^{1/2},
\label{1drot42}
\end{equation}
which is a basic relationship between thermodynamic quantities and the velocity
$c(T)$. Notice from (\ref{k5}) and (\ref{k7}) that to
leading logarithms $c(T) \approx c$, but this result is not satisfied by the
subleading terms. The characteristic excitation will have energy $\varepsilon \sim c
/\xi$, and this leads to our estimate for $\varepsilon$ made
earlier, when we justified the validity of a classical model.

The classical dynamics problem defined
by (\ref{1drot44}) and (\ref{1drot45a})
obeys that the crucial property of being free of all
ultraviolet divergences. Consequently,
we may determine its characteristic length and time scales by
simple engineering dimensional analysis, as no short distance
cutoff scale is going to transform into an anomalous dimension.
Indeed, a straightforward analysis shows that this classical
problem is free of dimensionless
parameters, and is a {\em unique}, parameter-free theory.  This is seen by defining
\begin{eqnarray}
 \overline{x} &=&  \frac{x}{ \xi} \nonumber \\
 \overline{t} &=&  \frac{t}{\tau_{\varphi}}  \nonumber \\
 \overline{{\bf L}} &=&  {\bf L}
\sqrt{\frac{\xi}{T \chi_{u\perp}}},
\label{1drot49}
\end{eqnarray}
where the characteristic time, $\tau_{\varphi}$,
is given by
\begin{equation}
\tau_{\varphi} = \sqrt{\frac{\xi  \chi_{u\perp}}{T}};
\label{1drot49a}
\end{equation}
our notation suggests that $\tau_{\varphi}$ (like $\Gamma_R^{-1}$
earlier) is a phase coherence time beyond which relaxational
dynamics of damped spin waves takes over.
Inserting (\ref{1drot49},\ref{1drot49a}) into (\ref{1drot44}) and (\ref{1drot45a}), we
find that all parameters disappear and the partition function and
equations of motion acquire a unique, dimensionless form, given by
setting $T=\xi=\chi_{u \perp}=1$ in them.

\begin{figure}
\epsfxsize=4in
\centerline{\epsffile{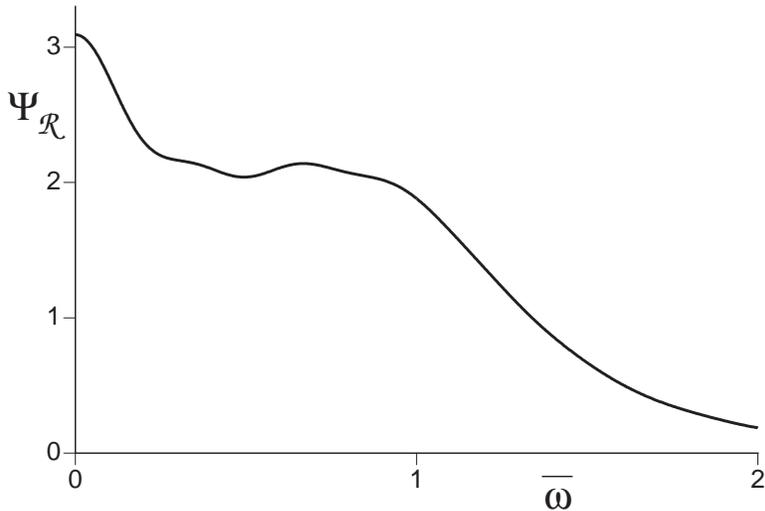}}
\caption{
Numerical results \protect\cite{chiran} for the scaling
function, $\Psi_{{\cal R}} ( 0, \overline{\omega})$, for the relaxation
function ${\cal R}$ of the $d=1$, $n=3$ case of ${\cal Z}_Q$ appearing in
(\protect\ref{1drotss}). The dimensionless frequency
$\overline{\omega} = \omega \tau_{\varphi}$.
}
\label{koreaf5}
\end{figure}

The above transformations allow us to easily obtain a scaling
form for the relaxation function ${\cal R}$:
\begin{equation}
{\cal R} (k, \omega) = \tau_{\varphi} \Psi_{{\cal R}} (k \xi, \omega \tau_{\varphi} ),
\label{1drotss}
\end{equation}
where $\Psi_{{\cal R}}$ is a universal scaling function, normalized as in
(\ref{k2}).
Further information on
the structure of $\Psi_{{\cal R}}$ was obtained \cite{chiran} by
a combination of analytic and
numerical methods.
At sufficiently large $k \xi$, we expect a pair of broadened, reactive,
`spin-wave' peaks at $\omega \approx c(T) k$ (with $c(T)$ given in
(\ref{1drot42})), which are similar to those found in the high $T$
limit of the quantum Ising chain in Fig~\ref{koreaf4}.
For the opposite limit of small $k\xi$, we present numerical
results for $\Psi_{{\cal R}} (0, \overline{\omega})$
in Fig~\ref{koreaf5}.
There is a sharp relaxational peak at
$\overline{\omega}=0$,
which is again similar to that found in the high $T$ limit of the
quantum Ising chain in Fig~\ref{koreaf4}. However, there is now a
well-defined shoulder at $\overline{\omega} \approx 0.7$
which was not found in the Ising case. This shoulder is a remnant
of the large $n$ result \cite{CSY,joli} which predicts a delta
function at $\omega \sim T/\ln(T/\Delta)$. So $N=3$ is large enough for this
finite frequency oscillation to survive in the high $T$ limit.

There is alternative, helpful way to view this oscillation frequency.
The underlying degree of freedom in our dynamical field theory
has a fixed amplitude, with $|{\bf n}| = 1$. However,
correlations of ${\bf n}$ decay
exponentially on a length scale $\xi$---so if we imagine coarse-graining
out to $\xi$, it is reasonable to expect significant {\em amplitude
fluctuations}
in the coarse-grained field. It is now useful
to visualize an effective field $\phi_{\alpha}$ with no length
constraint, which is just the field we introduced in Section~\ref{intro}.
On a length scale of order $\xi$,
we expect the effective potential controlling fluctuations of
$\phi_{\alpha}$ to have minimum at a non-zero value of
$|\phi_{\alpha}|$, but to also allow fluctuations in $|\phi_{\alpha}|$
about this minimum.
The finite frequency in Fig~\ref{koreaf5} is due to the harmonic
oscillations of $\phi_{\alpha}$ about this potential minimum,
while the dominant peak at $\omega =0$ is due to angular
fluctuations along the zero energy
contour in the effective potential.
This is interpretation is also consistent with the large $n$
limit, in which we freely integrate over all components of ${\bf n}$,
and so angular and amplitude fluctuations are not
distinguished.  The above argument could also have been applied to
the quantum Ising chain (in this case, angular fluctuations are replaced by
low-energy domain wall motion),
but the absence of such a reactive, finite frequency peak
at $k=0$ in Fig~\ref{koreaf4} indicates that $n=1$ is too far
from $n=\infty$ for any remnant of this large $n$ physics to
survive.

\section{Order parameter dynamics in $d=2$}
\label{sec:eps}

Determination of the long time dynamics in $d=2$ is a strongly
coupled problem and remains unsolved. As in the analysis in
Section~\ref{sec:ising},
we expect a phase coherence
time, $\tau_{\phi}$, and an inverse relaxation rate, $\Gamma_R^{-1}$,
of order $T$, and quantum and thermal fluctuations to play an
equally important role.

However, it has been argued recently \cite{ssrelax}, that within
the context of an expansion in
\begin{equation}
\epsilon = 3-d,
\label{k8}
\end{equation}
it is possible to separate quantum and thermal effects into two
distinct stages of the calculation, and to derive an effective
{\em classical wave model} for the long time dynamics. As in
Section~\ref{sec:1drot}, we  first derive an effective action for
the static susceptibility, and then supplement it with
equations of motion to obtain the unequal time correlations.
We define the zero Matsubara frequency component of $\phi_{\alpha}$
by
\begin{equation}
\Phi_{\alpha} (x) = T \int_0^{1/T} d \tau \, \phi_{\alpha} (x, \tau),
\label{k9}
\end{equation}
and derive an effective action for $\Phi_{\alpha}$ by
integrating out the non-zero Matsubara frequency components of
$\phi_{\alpha}$. To leading non-trivial order in an expansion in
$\epsilon$,
this leads to the following effective action
\begin{eqnarray}
&& {\cal Z}_{2C} = \int {\cal D} \Phi_{\alpha} (x) {\cal D} \Pi_{\alpha} (x)
\exp \left( - \frac{{\cal H}_{2C}}{T}
\right) \nonumber \\
&& {\cal H}_{2C} =  \int d^d x \, \left\{
\frac{1}{2}\left[ c^2 \Pi_{\alpha}^2 +
(\nabla_x \Phi_{\alpha})^2 +
\widetilde{R} \, \Phi_{\alpha}^2 \right]
+ \frac{U}{4!} \left( \Phi_{\alpha}^2 \right)^2 \right\} .
\label{calc}
\end{eqnarray}
As in (\ref{1drot44}), along with the functional integral over $\Phi_{\alpha} (x)$,
we have included an integral over a conjugate momentum field $\Pi_{\alpha} (x)$
which will be important for our subsequent treatment of the
dynamics; for now it easy to see that the Gaussian integral over
$\Pi_{\alpha}$ can be performed exactly, and it leaves the
correlations of $\Phi_{\alpha}$ under ${\cal Z}_{2C}$ unchanged. The
coupling constants in (\ref{calc}) are universally related to the
underlying field theory ${\cal Z}_Q$
controlling the quantum critical point. Before specifying these,
it is crucial to understand the nature of the ultraviolet
divergences in ${\cal Z}_{2C}$ considered as a classical field theory
in its own right. From standard field-theoretic analyses
\cite{ramond} it is known that ${\cal Z}_{2C}$ has only {\em one}
ultraviolet divergence, coming from a single one-loop tadpole
graph in the self energy: consequently, by trading the bare `mass'
$\widetilde{R}$ for a renormalized mass $R$ defined by
\begin{equation}
\widetilde{R} = R - T U \left(\frac{n+2}{6} \right) \int^{\Lambda}
\frac{d^d k}{(2 \pi)^d} \frac{1}{k^2 + R},
\label{defR}
\end{equation}
we can remove all cutoff dependencies in the correlators of ${\cal
Z}_{2C}$ (there are some additional divergences, associated with composite
operators,
which appear when two or more field operators approach each other in space: we
will not be concerned with these here).
All observables of ${\cal Z}_{2C}$ are then universal
functions of the couplings $R$ and $U$. Moreover, it is precisely these
couplings that are universally computed from the underlying
quantum field theory ${\cal Z}_Q$. Actually, instead to dealing
with $R$ and $U$ as the two independent parameters controlling
correlators of ${\cal Z}_{2C}$, it is convenient to replace $U$ by
the dimensionless parameter ${\cal G}$ defined by
\begin{equation}
{\cal G} \equiv \frac{TU}{R^{(4-d)/2}},
\label{k10}
\end{equation}
which is analogous to the Ginzburg parameter.
In the high $T$ limit (\ref{k1}), these parameters were shown to
have the following \cite{sseps,ssrelax} universal values to leading order in $\epsilon$
\begin{eqnarray}
R &=& \epsilon \left( \frac{n+2}{n+8} \right) \frac{2 \pi^2
(T/c)^2}{3} \nonumber \\
{\cal G} &=& \sqrt{\epsilon} \frac{48 \pi \sqrt{3}}{\sqrt{2 (n+2)(n+8)}}.
\label{rest}
\end{eqnarray}
As one lowers the temperature from the continuum high $T$ limit,
both $R$ and ${\cal G}$ vary as universal functions of $r/T^{1/(z
\nu)}$: $R$ decreases and ${\cal G}$ increases as we lower the
temperature into the magnetically ordered region ($r<0$), while
$R$ increases and ${\cal G}$ decreases as we lower the
temperature into the quantum paramagnetic region ($r>0$).

The values in (\ref{rest}) are the key to the argument justifying
the use of a classical dynamical model for small $\epsilon$. From
(\ref{calc}) it is clear that the characteristic $\Phi_{\alpha}$
fluctuations have an energy of order $c \sqrt{R} \sim \sqrt{\epsilon}
T$. As in Section~\ref{sec:1drot}, this energy is parametrically
smaller than $T$, and so the occupation number of the relevant $\Phi_{\alpha}$
will be given by their classical equipartition value.
This also means that the classical fluctuation dissipation theorem
is obeyed, and the static susceptibility, $\chi (k)$, computed
from (\ref{calc}) by
\begin{equation}
T \chi (k) = \frac{1}{n} \sum_{\alpha=1}^{n} \langle |
\Phi_{\alpha} (k) |^2 \rangle,
\label{k11}
\end{equation}
is also the equal-time correlation of $\phi_{\alpha}$.

We can now specify the recipe to compute the unequal time correlations
of $\phi_{\alpha}$, in manner which parallels
Section~\ref{sec:1drot}. The fundamental Poisson bracket is
\begin{equation}
\{\Phi_{\alpha} (x), \Pi_{\beta} (x') \}_{PB} = \delta_{\alpha
\beta} \delta(x-x'),
\label{k12}
\end{equation}
and the Hamiltonian ${\cal H}_{2C}$ then leads to the Hamilton-Jacobi
equations of motion
\begin{eqnarray}
\frac{\partial \Phi_{\alpha}}{\partial t} &=&
\left\{ \Phi_{\alpha} (x) , {\cal H}_{2C} \right\}_{PB} \nonumber \\
&=& c^2 \Pi_{\alpha},
\label{r6}
\end{eqnarray}
and
\begin{eqnarray}
\frac{\partial \Pi_{\alpha}}{\partial t} &=&
\left\{ \Pi_{\alpha} (x) , {\cal H}_{2C} \right\}_{PB} \nonumber \\
&=&  \nabla_x^2 \Phi_{\alpha} - \widetilde{R}
\Phi_{\alpha} - \frac{U}{6} (\Phi_{\beta}^2) \Phi_{\alpha},
\label{r7}
\end{eqnarray}
The equations (\ref{calc}), (\ref{r6}) and (\ref{r7}) define the
central dynamical non-linear wave model of this section. We will compute
correlations of the field $\Phi_{\alpha}$ at
unequal times, averaged over the set of initial conditions
specified by (\ref{calc}).
Notice all the thermal `noise' arises only in the random set of
initial conditions. The subsequent time evolution is then
completely deterministic, and precisely conserves energy,
momentum, and total ${\rm O} (n)$ charge. This should be
contrasted with the classical dynamical models studied in the
theory of dynamic critical phenomena~\cite{hhm,halphoh}, where there
are explicit damping co-efficients, along with
statistical noise terms, in the equations of motion.

The dynamical model has been defined above in the continuum,
and so we need to consider the nature of its short distance
singularities. As in Section~\ref{sec:1drot},
we assert \cite{ssrelax} that the {\em only\/}
short distance singularities are those already present in the
equal time correlations analyzed earlier.
These were removed by the simple renormalization in (\ref{defR}),
which is therefore adequate also for the unequal time
correlations.
With this knowledge in hand, we can immediately write down the
universal scaling form obeyed by the relaxation function by simple
arguments based upon analysis of engineering dimensions.
The analog of (\ref{1drotss}) is now
\begin{equation}
{\cal R} (k, \omega) = \frac{1}{c \sqrt{R}} \Psi_{{\cal R}} \left(
\frac{k}{\sqrt{R}}, \frac{\omega}{c \sqrt{R}}, {\cal G} \right),
\label{k13}
\end{equation}
where $\Psi_{{\cal R}}$ is a universal function we wish to
determine.

It now remains to solve the dynamical problem specified by
(\ref{calc}), (\ref{r6}) and (\ref{r7}), and so determine $\Psi_{{\cal
R}}$.
For small $\epsilon$, the dimensionless strength of the
non-linearity ${\cal G}$ in (\ref{rest}) is small; nevertheless we
cannot use perturbation theory in ${\cal G}$, because this fails
in the low frequency limit. In other words, the dynamical problem
remains strongly coupled even for small $\epsilon$.

The only remaining possibility is to numerically solve the strong-coupling
dynamical problem.
Formally, we are carrying out an $\epsilon$ expansion, and so the
numerical solution should be obtained for $d$ just below 3.
However, it is naturally much simpler to simulate {\em
directly in $d=2$}, which is also the dimensionality of physical interest.
Therefore, the approach to the solution of the dynamic
problem in the quantum critical region breaks down into two
systematic steps: ({\em i}) Use the $\epsilon=3-d$ expansion to
derive an effective classical non-linear wave problem \cite{sseps} characterized by the
couplings $R$ and ${\cal G}$. ({\em ii}) Obtain the exact
numerical solution \cite{ssrelax} of the classical non-linear wave problem at
these values of $R$ and ${\cal G}$ directly in $d=2$. This
division of the problem into two rather disjointed steps is also
physically reasonable: it is primarily for the classical thermal
fluctuations that the dimensionality $d=2$ plays a special
role, and the cases $n=1,2$ (which have non-zero temperature phase transitions
above the magnetically ordered phase) and $n \geq 3$ (which do not
have a non-zero temperature
phase transition) are strongly
distinguished---so it is important to treat these exactly; on
the other hand, the $\epsilon=3-d$ expansion provides a reasonable
treatment of the quantum fluctuations down to $d=2$ for all $n$.

\begin{figure}
\epsfxsize=3.7in
\centerline{\epsffile{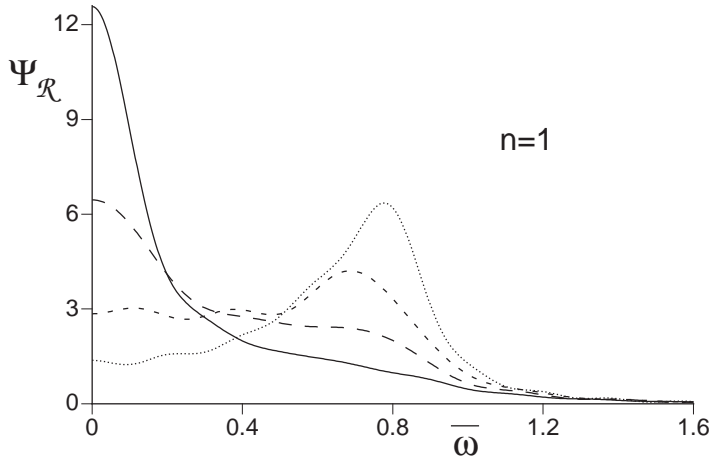}}
\caption{
Numerical results \protect\cite{ssrelax} for the zero momentum
scaling function $\Psi_{{\cal R}} (0, \overline{\omega}, {\cal G})$
for the relaxation function, ${\cal R}$, appearing in
(\protect\ref{k13}) for the $d=2$, $n=1$ model ${\cal Z}_Q$.
The dimensionless frequency $\overline{\omega} = \omega/c
\protect\sqrt{R}$.
Results are shown for
${\cal G} = 25$ (dots), ${\cal G} = 30$
(short dashes), ${\cal G} = 35$ (long dashes) and ${\cal G} = 40$
(full line). The high $T$ limit value of ${\cal G}$ in
(\protect\ref{rest}) evaluates to ${\cal G} = 35.5$ at
$\epsilon=1$ and $n=1$.
}
\label{koreaf6}
\end{figure}

\begin{figure}
\epsfxsize=3.7in
\centerline{\epsffile{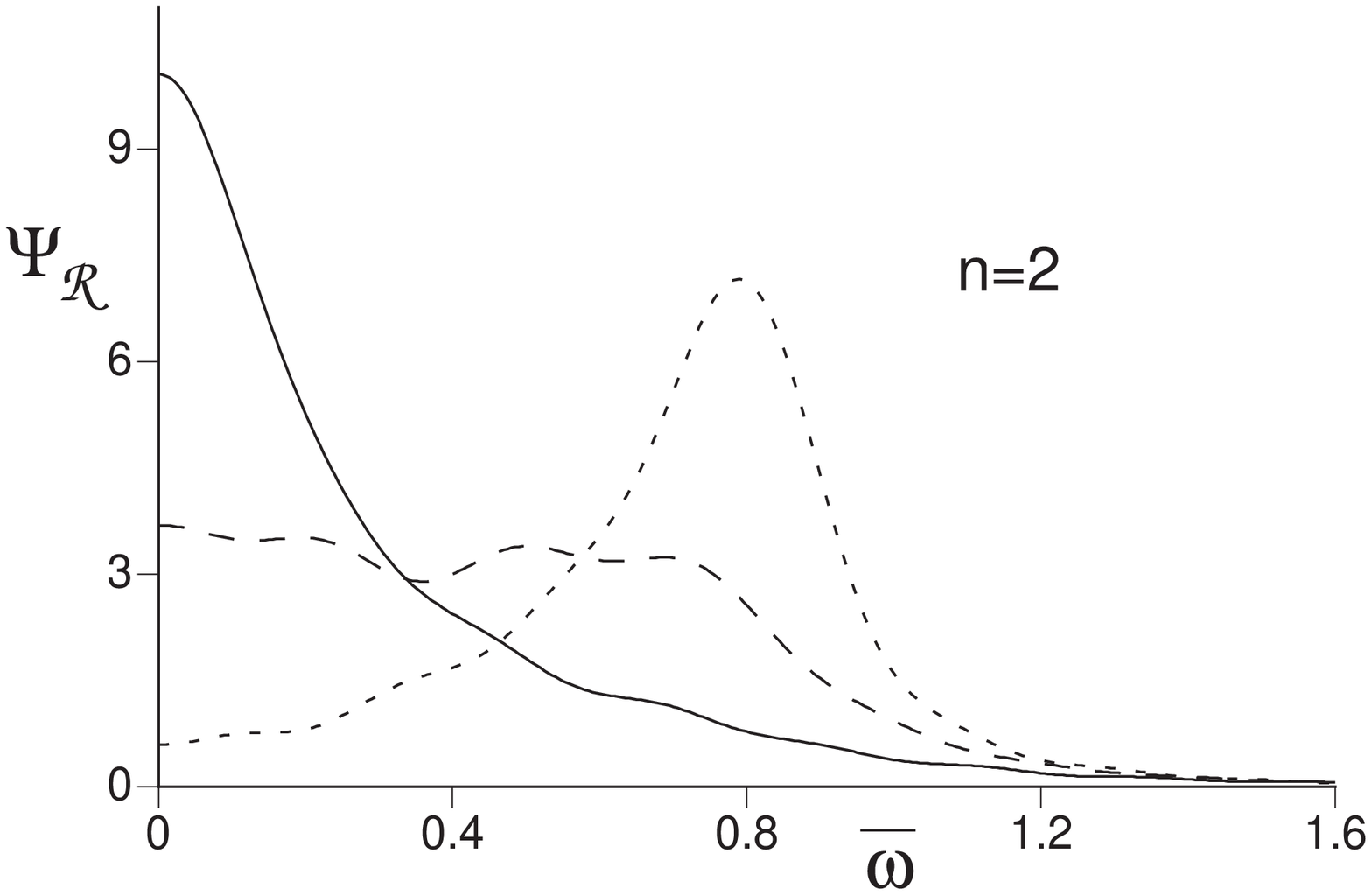}}
\caption{
As in Fig~\protect\ref{koreaf6} but for $d=2$, $n=2$. The values of ${\cal G}$
are now ${\cal G} = 20$
(short dashes), ${\cal G} = 30$ (long dashes) and ${\cal G} = 40$
(full line). The high $T$ limit value of ${\cal G}$ in
(\protect\ref{rest}) evaluates to ${\cal G} = 29.2$ at
$\epsilon=1$ and $n=2$.
}
\label{koreaf7}
\end{figure}

\begin{figure}
\epsfxsize=3.7in
\centerline{\epsffile{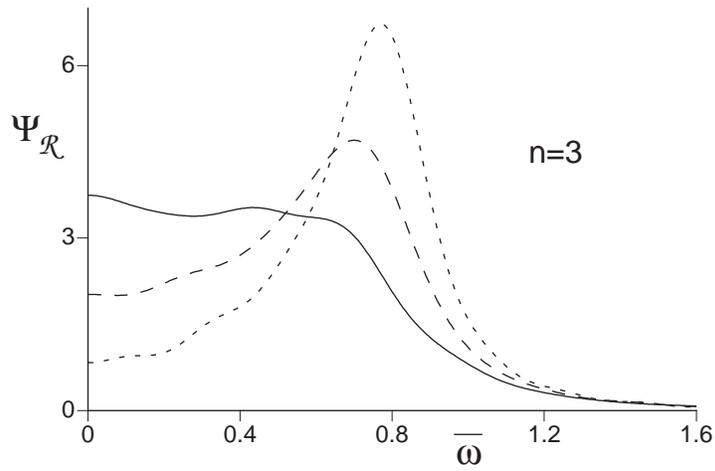}}
\caption{
As in Fig~\protect\ref{koreaf6} but for $d=2$, $n=3$. The values of ${\cal G}$
are now ${\cal G} = 20$
(short dashes), ${\cal G} = 25$ (long dashes) and ${\cal G} = 30$
(full line). The high $T$ limit value of ${\cal G}$ in
(\protect\ref{rest}) evaluates to ${\cal G} = 24.9$ at
$\epsilon=1$ and $n=3$.
}
\label{koreaf8}
\end{figure}

Figs~\ref{koreaf6}--\ref{koreaf8} contain
the results of a recent numerical computation of
the scaling function in (\ref{k13}) at $k=0$.
These results are the analog of
Fig~\ref{koreaf4} for the Ising chain and Fig~\ref{koreaf5} for
the $d=1$, $n=3$ case.
They show a consistent trend from small values of ${\cal G}$
and large values of $n$ to large values of ${\cal G}$ and small
values of $n$, and we discuss the physical interpretation of the
two limiting cases in turn.

For smaller ${\cal G}$ and larger $n$, we observe a
peak in ${\cal R}(0,\omega)$ at a non-zero frequency. This peak is the
remnant of a delta function obtained in the large $n$ limit
at a frequency $\omega \sim T$.
In the present computation, it is clear that the peak is due to
{\em amplitude fluctuations\/}
as $\Phi_{\alpha}$ oscillates about
the minimum in its effective potential at $\Phi_{\alpha} = 0$.
As ${\cal G}$ is reduced, we move out of the high $T$ region into
the low $T$ region above the quantum paramagnet,
and this finite frequency, amplitude fluctuation
peak connects smoothly with the quantum paramagnetic quasiparticle
peak. Of course, once we are in the quantum paramagnetic region,
the wave oscillations get quantized, and
the amplitude and width of the peak can no longer be computed
by the present quasi-classical {\em wave} description---we need an
approach which treats the excited {\em particles}
quasi-classically.

For larger ${\cal G}$ and smaller $n$, the peak in ${\cal R}(0, \omega)$
shifts down to $\omega = 0$. The resulting spectrum is then closer
to the exact solution for $d=1$, $n=1$ presented in
Fig~\ref{koreaf4}. As ${\cal G}$ increases further, the zero
frequency peak becomes narrower and taller.
How do we understand the dominance of this low frequency
relaxation~? For $n \geq 2$, there is a natural direction for
low energy motion of the order parameter: in angular or phase
fluctuations of
$\Phi_{\alpha}$
in a region where the value of $|\Phi_{\alpha}|$ is non-zero. Of
course, the {\em fully} renormalized effective potential controlling
fluctuations of $\Phi_{\alpha}$ has a minimum only at $\Phi_{\alpha} =
0$, as we are examining a region with no long range order.
However, for these values of ${\cal G}$,
there is a significant intermediate length scale
over which the local effective potential has a minimum at
a $|\Phi_{\alpha}| \neq 0$, and the predominant fluctuations
of $\Phi_{\alpha}$ consist of a relaxational phase dynamics.

The above reasoning has been for the cases with continuous
symmetry, $n \geq 2$. However, closely related arguments can also
be made for $n=1$. In this case, in a region where
$|\Phi_{\alpha}|$ locally takes a non-zero value, there are
low-energy modes corresponding to motions of {\em domain
walls} between oppositely oriented magnetic
phases. Indeed, precisely such a domain wall motion
was mentioned for the $d=1$, $n=1$ case, and was argued to be
behind the relaxational peak in Fig~\ref{koreaf4}.

Even in a region dominated by angular (or domain-wall)
fluctuations about a locally
non-zero value of $|\Phi_{\alpha}|$, there could still be higher
frequency amplitude fluctuations of $|\Phi_{\alpha}|$ about its
local potential minimum. This would be manifested by
peaks in ${\cal R}(0, \omega)$ both at $\omega = 0$ and at a non-zero
frequency. A glance at Figs~\ref{koreaf6}--\ref{koreaf8} shows that
this never happens in a well-defined manner. However, for $n=1$,
we do observe a non-zero frequency shoulder in ${\cal R}(0, \omega)$
at ${\cal G}=35$, along with a prominent peak at $\omega=0$: this
indicates the simultaneous presence of domain wall relaxational
dynamics and amplitude fluctuations in $|\Phi_{\alpha}|$.
Readers will also recognize the similarity
of this with the shoulder in
Fig~\ref{koreaf5} describing the high $T$ limit of the $d=1$, $n=3$
case.
For the other cases in $d=2$, we do not see a clear signal of the
concomitant amplitude and angular fluctuations:
it appears, therefore, that once angular fluctuations appear with
increasing ${\cal G}$, the non-linear couplings between
the modes reduce the spectral weight in the amplitude mode to
a negligible amount.

It is interesting to examine the above results at the value of
high $T$ limit for ${\cal G}$
in (\ref{rest}) evaluated directly in $\epsilon = 1$. We find
${\cal G} = 35.5$, $29.2$, $24.9$ for $n=1$, $2$, $3$, and these values are very
close to the position where the crossover between the above
behaviors occurs. The $n=1$ case has a clear maximum
in ${\cal R} (0, \omega)$ at $\omega =0$ (along with a finite frequency shoulder),
 while there is a more clearly
defined finite frequency peak for $n=3$.

In closing, we note that there is a passing
resemblance between
the above crossover in dynamical properties as a function of ${\cal
G}$,
and a well-studied phenomenon in dissipative
quantum mechanics \cite{leggett,dissqm,lesage}: the crossover from `coherent oscillation' to
`incoherent relaxation' in a two-level system
coupled to a heat bath . However, here we do not rely on an
arbitrary heat bath of linear oscillators, and the relaxational
dynamics emerges on its own from the underlying
Hamiltonian dynamics of an interacting many-body, quantum system.
Our description of the crossover has been carried out in
the context of a quasi-classical wave model here, but, as we noted earlier,
the `coherent' peak
connects smoothly to the quasiparticle peak in low $T$
paramagnetic region---here the wave oscillations get
quantized into discrete lumps which must then be described by
a `dual' quasi-classical particle picture.

\section{Conclusions}
\label{sec:conc}
We have described the high temperature relaxational dynamics for a
number of models in spatial dimensions $d=1,2$. This dynamics is
a property of a renormalizable, interacting continuum quantum field
theory. Two cases can be further distinguished:
\newline
({\em i}) The
excitations of the theory retain a non-zero scattering amplitude
at high energies and temperatures: the models of
Section~\ref{sec:ising} and~\ref{sec:eps} are of this type. For
these, the only characteristic energy scale controlling the
density and interaction strength of the excitations becomes
$T$ itself, and so the phase coherence time, and the inverse
relaxation rate, are universal numbers times $\hbar/k_B T$.
As a result, quantum and thermal fluctuations contribute
equally to the phase relaxation. (However, in Section~\ref{sec:eps}
we did develop an expansion in which the universal prefactors
of $\hbar/k_B T$ became numerically large and so the long time relaxation was
described by an effective classical model.)
\newline
({\em ii}) The theory becomes asymptotically free at high
energies, and so the scattering amplitude of the excitations
vanishes at large $T$. The model of Section~\ref{sec:1drot} is
of this class, and has a phase coherence time and inverse
relaxation rate of order $(\hbar/k_B T) \ln (k_B T/\Delta)$,
where $\Delta$ is an energy scale characterizing the low
energy theory. These times are parametrically larger than
$\hbar /k_B T$ and so the relaxational dynamics is classical.

We have developed a fairly complete description of the dynamical
correlations of these models, and our results should be testable
in experiments on compounds the cuprate
superconductor family, Heisenberg spin chains, and double layer
quantum Hall systems \cite{pellegrini,DSZ}.

\section*{Acknowledgments}
The results in Section~\ref{sec:1drot} grew out of collaborations
with Kedar Damle~\cite{dslong} and Chiranjeeb
Buragohain~\cite{chiran}.

Portions of this review have been adapted from
``Quantum Phase Transitions'', by S.~Sachdev,
Cambridge University Press, in press. I am grateful to the
Press for permission to use this material here.

I thank Professors Yunkyu~Bang, Y.~M.~Cho, Jisoon~Ihm, Jaejun~Yu
and Lu~Yu for the opportunity to attend this stimulating
conference, and for their hard work in making it a great success.
This research was supported by NSF Grant No DMR 96--23181.


\section*{References}
\begin{thebibliography}{99}

\bibitem{aeppliscience} G.~Aeppli, T.~E.~Mason, S.~M.~Hayden,
H.~A.~Mook, and J.~Kulda, \Journal{\SCN}{278}{1432}{1998}.

\bibitem{SY} S.~Sachdev, and J.~Ye, \Journal{\PRL}{69}{2411}{1992}.

\bibitem{CSY} A.~V.~Chubukov, S.~Sachdev, and J.~Ye
\Journal{\PRB}{49}{11919}{1994}.

\bibitem{tran} J.~M.~Tranquada, J.~D.~Axe, N.~Ichikawa, A.~R.~Moodenbaugh,
Y.~Nakamura and S.~Uchida \Journal{\PRL}{78}{338}{1997}.

\bibitem{CHN} S.~Chakravarty, B.~I.~Halperin, and D.~R.~Nelson,
\Journal{\PRB}{39}{2344}{1989}.

\bibitem{statphys} S. Sachdev in {\em Proceedings of the 19th IUPAP
International Conference on Statistical Physics, Xiamen, China\/},
ed. B.-L. Hao, (World Scientific, Singapore, 1996); cond-mat/9508080.

\bibitem{apy} S. Sachdev and A.~P.~Young,
\Journal{\PRL}{78}{2220}{1997}.

\bibitem{dslong} K.~Damle and S.~Sachdev, \Journal{\PRB}{57}{8307}{1998}.

\bibitem{chiran} C.~Buragohain and S.~Sachdev, cond-mat/9811083.

\bibitem{joli} Th.~Jolicoeur and O.~Golinelli, {\it Phys. Rev. B}
{\bf 50}, 9265 (1994).

\bibitem{ssrelax} S.~Sachdev, cond-mat/9810399.

\bibitem{ramond} P.~Ramond, {\it Field Theory, A Modern Primer}
(Benjamin-Cummings, Reading, 1981).

\bibitem{sseps} S.~Sachdev, \Journal{\PRB}{55}{142}{1997}.

\bibitem{hhm} B.~I.~Halperin, P.~C.~Hohenberg, and S.~k.~Ma,
{\it Phys. Rev. Lett.} {\bf 29}, 1548 (1972);
{\it Phys. Rev. B} {\bf 10}, 139 (1974).

\bibitem{halphoh} P.~C.~Hohenberg and B.~I.~Halperin, {\it Rev. Mod.
Phys.}
{\bf 49}, 435 (1977).

\bibitem{pellegrini} V.~Pellegrini, A.~Pinczuk, B.~S.~Dennis, A.~S.~Plaut,
L.~N.~Pfeiffer, and K.~W.~West
{\it Science} {\bf 281}, 799 (1998).

\bibitem{DSZ} S.~Das Sarma, S.~Sachdev, and L.~Zheng,
{\it Phys. Rev. B} {\bf 58}, 4672 (1998).

\bibitem{leggett} A.~J.~Leggett, S.~Chakravarty, A.~T.~Dorsey, M.~P.~A.~Fisher,
A.~Garg, and W.~Zwerger, {\it Rev. Mod. Phys.} {\bf 59}, 1 (1987).

\bibitem{dissqm} U.~Weiss, {\em Quantum Dissipative Systems}
(World Scientific, Singapore, 1993).

\bibitem{lesage} F.~Lesage and H.~Saleur, {\it Nucl. Phys. B}
{\bf 493}, 613 (1997).


\end{thebibliography}
\end{document}